\documentclass[
reprint,
  nofootinbib,amsmath,amssymb,aps,preprintnumbers,superscriptaddress,notitlepage
]{revtex4-1}

\usepackage{graphicx}
\usepackage{bm}
\usepackage{braket}
\usepackage{cases} 
\usepackage{here}
\usepackage{color}
\usepackage{upgreek}
\usepackage{comment}
\usepackage{empheq}

\usepackage{hyperref}

\begin{document}

\preprint{KEK-QUP-2023-0018, KEK-TH-2558, KEK-Cosmo-0327, TU-1205}

\title{Gravitational wave search through electromagnetic telescopes}

\author{Asuka Ito}
\email[]{asuka.ito@kek.jp}
\affiliation{International Center for Quantum-field Measurement Systems for Studies of the Universe and Particles (QUP), KEK, Tsukuba 305-0801, Japan}
\affiliation{Theory Center, Institute of Particle and Nuclear Studies, KEK, Tsukuba 305-0801, Japan}

\author{Kazunori Kohri}
\email[]{kohri@post.kek.jp}
\affiliation{Division of Science, National Astronomical Observatory of Japan (NAOJ), and SOKENDAI, 2-21-1, Osawa, Mitaka, Tokyo 181-8588, Japan}
\affiliation{Theory Center, Institute of Particle and Nuclear Studies, KEK, Tsukuba 305-0801, Japan}
\affiliation{International Center for Quantum-field Measurement Systems for Studies of the Universe and Particles (QUP), KEK, Tsukuba 305-0801, Japan}
\affiliation{Kavli IPMU (WPI), UTIAS, The University of Tokyo, Kashiwa, Chiba 277-8583, Japan}

\author{Kazunori Nakayama}
\email[]{kazunori.nakayama.d3@tohoku.ac.jp}
\affiliation{Department of Physics, Tohoku University, Sendai, Miyagi 980-8578, Japan}
\affiliation{International Center for Quantum-field Measurement Systems for Studies of the Universe and Particles (QUP), KEK, Tsukuba 305-0801, Japan}
\affiliation{Kavli IPMU (WPI), UTIAS, The University of Tokyo, Kashiwa, Chiba 277-8583, Japan}


\begin{abstract}
We study the graviton-photon conversion in the magnetic fields of the Earth, the Milky Way Galaxy, and intergalactic regions.
Requiring that the photon flux converted from gravitons does not exceed the observed photon flux with telescopes,
we derive upper limits on the stochastic gravitational waves in frequency ranges from $10^{7}$Hz to $10^{35}$Hz.
Remarkably, the upper limits on $h^2 \Omega_{{\rm GW}}$ could be less than unity in the frequency range of $10^{18}$-$10^{23}$ Hz in a specific case.
The detection of gravitational waves using telescopes would open up a new avenue for 
high frequency gravitational wave observations.
\end{abstract}

\maketitle

%
%
%
%
%
%
%
\section{Introduction}
In 2015, LIGO/Virgo collaborations detected gravitational waves from a binary black hole merger for the first time~\cite{LIGOScientific:2016aoc}.
Recently, stochastic gravitational waves around nHz were detected with pulsar timing arrays~\cite{NANOGrav:2023gor,EPTA:2023fyk}.
Undoubtedly, it is important to push forward multi-frequency gravitational wave observations
in order to understand and investigate the evolution of our universe~\cite{Kuroda:2015owv}.
In the low frequency range around $10^{-18}-10^{-16}$\,Hz, a promising method for observing gravitational waves is to seek for the B-mode polarization of the cosmic microwave observations~\cite{Planck:2018jri,Paoletti:2022anb,LiteBIRD:2022cnt}.
On the other hand, detection of high frequency gravitational waves above kHz is still under development and 
even new ideas are required~\cite{Aggarwal:2020olq,Ito:2019wcb,Ito:2020wxi,Ito:2022rxn,Ejlli:2019bqj,Ringwald:2020ist,Aggarwal:2020olq,Berlin:2021txa,Domcke:2022rgu,Tobar:2022pie,Berlin:2023grv,Ito:2023bnu}, though such high frequency gravitational waves are theoretically
interesting as probes of new physics~\cite{Ghiglieri:2015nfa,Wang:2019kaf,Ghiglieri:2020mhm,Ringwald:2020ist,Ghiglieri:2022rfp,Vagnozzi:2022qmc,Ema:2021fdz,Ema:2015dka,Ema:2016hlw,Ema:2020ggo,Nakayama:2018ptw,Huang:2019lgd,Barman:2023ymn,Khlebnikov:1997di,Easther:2006gt,Easther:2006vd,Garcia-Bellido:2007nns,Ito:2016aai,Aggarwal:2020olq,Franciolini:2022htd,Saito:2021sgq,Gehrman:2022imk,Gehrman:2023esa}.

One natural approach for detecting high frequency gravitational waves is to employ tabletop experiments. 
This is because gravitational wave detectors tend to achieve high sensitivity when their size is on the same scale as the 
wavelength of the gravitational waves.
For example, the magnon gravitational wave detector utilizing resonant excitation of magnons by gravitational waves
was proposed for detecting gravitational waves around GHz~\cite{Ito:2019wcb,Ito:2020wxi,Ito:2022rxn}.
Under the presence of a background magnetic field, gravitational waves can also be converted into 
photons~\cite{gertsenshtein1962wave,Raffelt:1987im}. 
In this regard, there have been significant proposals for new high frequency gravitational wave detection methods utilizing axion detection experiments~\cite{Ejlli:2019bqj,Ringwald:2020ist,Aggarwal:2020olq,Berlin:2021txa,Domcke:2022rgu,Tobar:2022pie,Berlin:2023grv}.

Another possible way is to utilize astrophysical observations of photons with various frequencies.
References \cite{Pshirkov:2009sf, Dolgov:2012be, Domcke:2020yzq, Ramazanov:2023nxz} put forward the idea that the detection of microwave/X-ray/gamma-ray photons can impose limits on the presence of stochastic gravitational waves at their respective frequencies. The constraints depend on the magnitude of the primordial/Galactic magnetic field.%
\footnote{
    The cosmic background photon conversion into gravitons has also been studied in Refs.~\cite{Chen:1994ch,Cillis:1996qy,Chen:2013gva,Fujita:2020rdx}.}
Recently, it was also pointed out that the graviton-photon conversion happens in the
magnetosphere of a planet~\cite{Liu:2023mll} or a pulsar~\cite{Ito:2023fcr}.
In this paper, we further investigate the possibility of the gravitational wave detection with various telescopes.
We study the graviton-photon conversion in magnetic fields of the Earth, in the Milky Way Galaxy, and in intergalactic regions and 
calculate photon flux expected to be observed with telescopes.
Comparing it with the observed photon spectra with various astrophysical photon observations,
we show that the gravitational wave detection with telescopes is quite promising.
Therefore, the detection of gravitational waves using telescopes would open up a new avenue and 
advance multi-frequency gravitational wave observations in the high frequency range.
\section{Graviton-photon conversion}
%
We consider the action
\begin{equation}
    S = \int d^{4}x \sqrt{-g}\bigg[ \frac{M_{{\rm pl}}^{2}}{2}R
                           -\frac{1}{4} F_{\mu\nu}F^{\mu\nu} 
                          \bigg] ,  \label{ac}
\end{equation}
where $M_{{\rm pl}}$ represents the reduced Planck mass, $R$ is the Ricci scalar,
$g$ is the determinant of the metric $g_{\mu\nu}$.
The field strength of electromagnetic fields is defined by 
$F_{\mu\nu} = \partial_{\mu}\mathcal{A}_{\nu} - \partial_{\nu}\mathcal{A}_{\mu}$ where $\mathcal{A}_{\mu}$ is the vector potential.
We now expand the vector potential and the metric as
\begin{empheq}[left=\empheqlbrace]{align} 
  \mathcal{A}_{\mu}(x) &= \bar{A}_{\mu} + A_{\mu}(x) , \\
  \label{met033}
  g_{\mu\nu}(x) &= \eta_{\mu\nu} + \frac{2}{M_{{\rm pl}}} h_{\mu\nu}(x) .       
\end{empheq}
Here, $\bar{A}_{\mu}$ consists of background magnetic fields $\bar{B}^{i} = \epsilon^{ijk} \partial_{j} \bar{A}_{k}$.
$\eta_{\mu\nu}$ stands for the Minkowski metric, and 
$h_{\mu\nu}(x)$ is a traceless-transverse tensor representing
canonically normalized gravitational waves.
Below we take the gauge $A_0=0$ and only consider two transverse modes.
Let us consider gravitons and/or photons propagating along $z$-direction and the background magnetic field orthogonal to the propagation direction,\footnote{
    Electromagnetic wave generation from gravitational waves in vacuum without background magnetic field was studied in Ref.~\cite{Jones:2017dzt}.
} which is taken to be $y$-direction, $\bar{\bm{B}}=(0, \bar{B}, 0)$.
One can also choose the polarization bases for the vector and the tensor as
\begin{eqnarray}
  &e^{+}_{i} =  \begin{pmatrix}
                     1 \\
                     0 \\
                     0
                    \end{pmatrix} , \quad
  e^{\times}_{i} =  \begin{pmatrix}
                     0 \\
                     1 \\
                     0
                    \end{pmatrix} , \quad  \nonumber \\
  &\epsilon^{+}_{ij}= \frac{1}{\sqrt{2}} \begin{pmatrix}
              1 & 0 & 0\\
              0 & -1 & 0\\
              0 & 0 & 0
              \end{pmatrix} ,  \quad
  \epsilon^{\times}_{ij}= \frac{1}{\sqrt{2}} \begin{pmatrix}
              0 & 1 & 0\\
              1 & 0 & 0\\
              0 & 0 & 0
              \end{pmatrix} .   \label{basis}
\end{eqnarray}
Using the basis (\ref{basis}), the electromagnetic field and the gravitational wave can be expanded as follows:
\begin{empheq}[left=\empheqlbrace]{align} 
  A_{i} = e^{-i(\omega t-kz)} A^{\sigma}(z) e^{\sigma}_{i} , \\
  h_{ij} = e^{-i(\omega t-kz)} h^{\sigma}(z) \epsilon_{ij}^{\sigma} .     \label{planar}
\end{empheq}

We can now derive coupled equations of motion for the photon 
and the graviton of each polarization modes from Eqs.\,(\ref{ac})-(\ref{planar}).
We also take into account the plasma effect and the vacuum polarization~\cite{2006physics...5038H}. 
As for the vacuum polarization effect, we consider corrections to the dispersion relation from both the background magnetic field and the cosmic microwave background (CMB)~\cite{Latorre:1994cv,Kong:1998ic,Thoma:2000fd,Dobrynina:2014qba}. 
The derived equations are
\begin{equation}
  \left[ i \partial_{z} 
         +
                 \begin{pmatrix}
              - \frac{\omega_{p}^{2}}{2\omega}  
              + \frac{\omega^{2}_{{\rm QED},\sigma}}{2\omega} 
              + \frac{\omega^{2}_{{\rm CMB}}}{2\omega}
              & i \frac{B}{\sqrt{2}M_{{\rm pl}}} \\
              -i \frac{B}{\sqrt{2}M_{{\rm pl}}} 
              & 0
              \end{pmatrix}
      \right] 
      \begin{pmatrix}
                     A^{\sigma}(z) \\
                     h^{\sigma}(z) 
      \end{pmatrix}
            \simeq 0 ,   \label{mat}
\end{equation}
with
\begin{eqnarray}
  \omega_{p} &=& \sqrt{ \frac{4 \pi \alpha n_{e}}{m_{e}} } , \nonumber \\ 
  \omega_{{\rm QED},\sigma} &=& \sqrt{  \frac{8\lambda_{\sigma}\alpha^{2}\omega^{2}B^{2}}{45m_{e}^{4}} },
  \nonumber \\
  \omega_{{\rm CMB}} &=& \sqrt{  \frac{88\pi^{2}\alpha^{2}\omega^{2}T^{4}}{2025 m_{e}^{4}} }  .         \label{omega}            
\end{eqnarray}
$\lambda_{\sigma}$ takes the value of 2 (7/2) for $\sigma=+$ ($\sigma=\times$), $\alpha$ is the fine structure constant, 
$T$ represents the temperature of the CMB,
 $n_{e}$ is the electron number density, and $m_{e}$ is the electron mass.
As deriving the equations,
we have assumed that the scale of conversion between photons and gravitons is much longer than $k^{-1}$ and photons
are ultrarelativistic, i.e., $\omega \simeq k$.
Also, we neglected spatial derivative of $B$.
From Eq.\,(\ref{omega}), we see that while $\frac{\omega_{p}^{2}}{2\omega}$ becomes significant in lower frequency regime,
$\frac{\omega_{{\rm QED},\sigma}^2}{2\omega}$ or $\frac{\omega_{{\rm CMB}}^2}{2\omega}$ gets significant in higher frequency regime in general.
In the next section, we will give upper limits on stochastic gravitational waves converted to photons in various 
magnetic fields in the universe with the use of telescope observations.
\section{Upper limits on stochastic gravitational waves}
In this section, we consider the graviton-photon conversion 
in the geomagnetic fields, in the Milky Way Galaxy, and in intergalactic regions.
We then calculate the photon flux converted from stochastic gravitational waves and 
compare it with the observed data of photon flux by various telescope in order to give upper limits on the abundance of 
stochastic gravitational waves.
%
%
\subsection{Graviton to photon conversion in the Earth's magnetic fields}\label{secearth}
It is known that the geomagnetic field envelops the Earth, extending from the magnetic North Pole to the magnetic South Pole.
The magnitude of the geomagnetism on the surface of the Earth is $B_{E} = 0.45$G on average~\cite{Alken2021}.
Since the geomagnetic field has a dipole like structure, the amplitude of the magnetic field scales as
\begin{equation}
  B(r) = B_{E}\left( \frac{r_{E}}{r} \right)^{3} ,  \label{dipo}
\end{equation}
where $r_{E} = 6367$km is the averaged radius of the Earth.
There is no plasma in the troposphere ($0\sim 10$\,km) and the stratosphere ($10\sim 50$\,km).
In the ionosphere ($50\sim 1000$\,km), there exists plasma whose density ranges from 
$n_{e,1}=10^{2}{\rm cm}^{-3}$ to $10^{6}{\rm cm}^{-3}$~\cite{Bilitza2011,bilitza}.
In the inner magnetosphere ($1000\sim 60000$\,km), there also exists plasma with the number density
$n_{e,2}\sim 10-10^{4}{\rm cm}^{-3}$~\cite{aaa,bbb}%
\footnote{In~\cite{Liu:2023mll}, which also studied graviton-photon conversion in the geomagnetic fields, 
the plasma effect is assumed to be negligible even above the altitude of $50$km.
}.
Outside the inner magnetosphere, it is not obvious how the magnetic field and the plasma density distribute due to 
substantial effects of solar wind.
Therefore, we will only consider the graviton-photon conversion within the altitude of 60000km to obtain 
a conservative result.
Photon flux converted from stochastic gravitational waves in the geomagnetic field can be calculated by solving
Eq.\,(\ref{mat}) with Eq.\,(\ref{dipo}) numerically.
We then give upper bounds on the stochastic gravitational waves by comparing the predicted photon flux with 
the observational results by various telescopes;
the CMB~\cite{Hill:2018trh},
the cosmic photon background (CPB)~\cite{Hill:2018trh},
active galactic nuclei~\cite{MAGIC:2022piy}, 
the Crab nebula~\cite{LHAASO:2021cbz}, 
and ultra high energy photons~\cite{PierreAuger:2022gkb}.

\begin{widetext}
\begin{figure}[H]
\centering
\includegraphics[width=16.5cm]{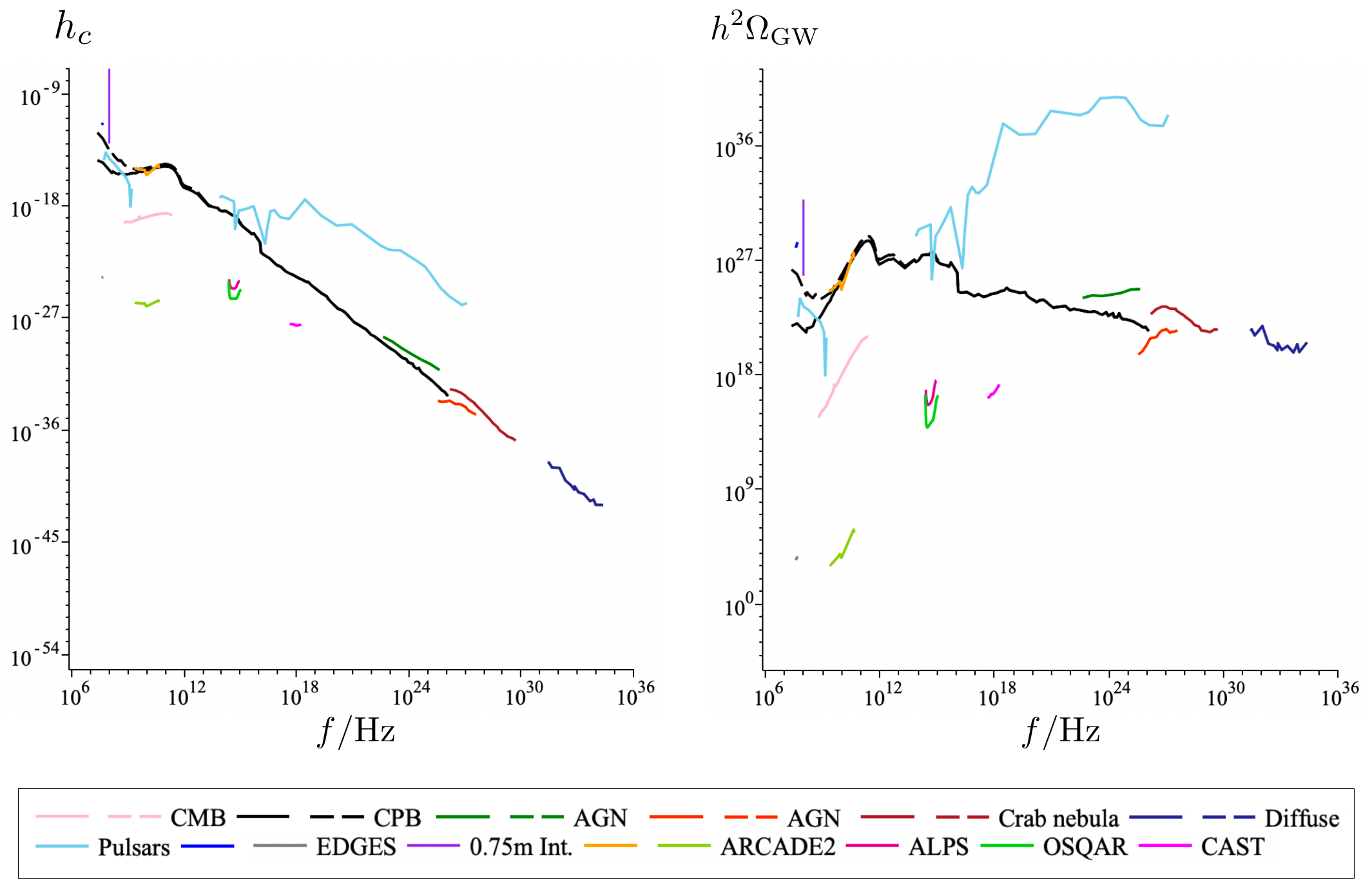}
\caption{Limits on stochastic gravitational waves converted to photons in the geomagnetic field with various telescope observations are shown.
The pink, black, deep green, red, brawn, and navy lines (dashed lines) are obtained from 
observations of the CMB~\cite{Hill:2018trh}, the CPB (COBE, AKARI, CIBER, HST, EUVE, MAXI, COMPTEL, FERMI)~\cite{Hill:2018trh},
active galactic nuclei of FERMI~\cite{MAGIC:2022piy} and of MAGIC~\cite{MAGIC:2022piy}, 
the Crab nebula (LHASSO)~\cite{LHAASO:2021cbz}
and ultra high energy photons (Pierre Auger)~\cite{PierreAuger:2022gkb}, respectively, for 
$n_{e,1}=10^{2}{\rm cm}^{-3}, n_{e,2}=10{\rm cm}^{-3}$ ($n_{e,1}=10^{6}{\rm cm}^{-3}, n_{e,2}=10^{4}{\rm cm}^{-3}$).
The light blue line represents limits from pulasr observations~\cite{Ito:2023fcr}.
The blue (grey) and the orange (lime green) lines respectively represents 
the constraints with EDGES and ARCADE2 for maximal (minimum) amplitude of cosmological magnetic fields~\cite{Domcke:2020yzq}.
The violet line is the upper limit from 0.75~m interferometer~\cite{Akutsu:2008qv}.
The wine red, green, and magenta lines represents constraints with ALPS, OSQAR, and CAST, respectively~\cite{Ejlli:2019bqj}.
} \label{earth}
\end{figure}
\end{widetext}

The result is shown in Fig.\,\ref{earth} where limits on
the characteristic amplitude $h_c$ defined by
$ \langle h_{ij}h^{ij}\rangle=M_{{\rm pl}}^2 /2 \times \int d(\ln f) h^{2}_{c}(f)$ and the energy density parameter, which 
is related to $h_c$ and the Hubble constant $H_0$ ($= h\times 100\,$km/s\,Mpc) as 
$\Omega_{\rm GW} = 2\pi^2 f^2 h_c^2/3H_0^2$, are depicted.
Note that the upper bound on $\Omega_{\rm GW}$ derived in this paper is much larger than unity in the most cases. Such a large value of $\Omega_{\rm GW}$ could still be meaningful if it is interpreted as the local quantity rather than the average over the whole universe.
The pink, black, deep green, red, brawn, and navy lines (dashed lines) are obtained from 
observations of the CMB on the ground~\cite{Hill:2018trh}, the CPB in space~\cite{Hill:2018trh} 
(COBE: 900km~\cite{cobe} , AKARI: 750km~\cite{akari}, CIBER: 577km~\cite{Matsuura:2017lub}, 
HST: 579km~\cite{hst}, 
EUVE: 527km~\cite{euve}, MAXI: 400km~\cite{maxi}, COMPTEL: 500km~\cite{comptel}, 
FERMI: 535km~\cite{fermi}),
active galactic nuclei~\cite{MAGIC:2022piy} (FERMI: 535km~\cite{fermi}, MAGIC: on the ground), 
the Crab nebula~\cite{LHAASO:2021cbz} (LHASSO: on the ground)
and ultra high energy photons~\cite{PierreAuger:2022gkb} (Pierre Auger: on the ground), respectively, for 
$n_{e,1}=10^{2}{\rm cm}^{-3}$ and $n_{e,2}=10{\rm cm}^{-3}$ ($n_{e,1}=10^{6}{\rm cm}^{-3}$ and $n_{e,2}=10^{4}{\rm cm}^{-3}$).
Field of view angle is taken as $0.25^\circ$ for FERMI~\cite{MAGIC:2022piy},
$1.5^\circ$ for MAGIC~\cite{MAGIC:2022acl},
$1^\circ$ for LHASSO~\cite{LHAASO:2021cbz}, 
and (approximately) $360^\circ$ for other telescopes.
In the lower frequency range where the plasma effect, which prevents the conversion, becomes important, 
the constraints depend on the plasma density.
Indeed, the result depends on the height from the ground of each detector because
the graviton-photon conversion efficiently occurs in the troposphere and the stratosphere for 
gravitational waves in the lower frequency range.
This is also the reason why the pink line and dashed line 
are almost indistinguishable.
Furthermore, in the frequency range higher than $\gtrsim 10^{13}$Hz,
the solid and dashed black, deep green, red, brawn, and navy lines are almost degenerates 
since the plasma effect is not significant in the frequency range.

It should be noted that we focused on photon flux due to graviton-photon conversion from the direction of space
even for satellite telescopes.
Then,
one could think that the conversion rate of gravitational waves to electromagnetic waves is parameterized by altitude alone, approximately.
Instead, one can consider photon flux due to graviton-photon conversion from the direction of Earth,
which allows us to block background photons, as is done in~\cite{Liu:2023mll}.
In such a case, the photon flux can be more anisotropic and sensitive to 
the detector's field and direction of view.
However, in our case, the direction dependence of photon flux is less important and can be neglected. 

One might consider that the attenuation of photons by the air should be taken into account for telescopes on the ground when considering the graviton-photon conversion in the geomagnetic field. 
In Fig.\,\ref{earth}, there are three types of experiments conducted on the ground: 
the CMB experiment represented by the pink line, MAGIC, and LHASSO. 
For the frequency range of the CMB experiment, the attenuation of photons by the air is small and negligible. 
For MAGIC and LHASSO, the attenuation effect is utilized for the detection of gamma-rays as air showers. 
Therefore, the attenuation effect in the air is not relevant to our discussion. 
\subsection{Graviton to photon conversion in the Milky Way Galaxy} \label{gala}
There exists magnetic fields in our galaxy~\cite{Haverkorn:2014jka,Boulanger:2018zrk}.
The typical strength of the magnetic field $B_{{\rm G}}$ ranges from $1\mu$G to $10\mu$G.
There are two kind of main magnetic field components; one is large scale magnetic fields, which have 
large coherence and directional dependence along the spiral of the galaxy and 
the other is small scale magnetic fields whose coherence length is about $1$pc 
$\sim 100$pc~\cite{Haverkorn:2004fw,2013A&A...558A..72I,Haverkorn:2008tb}.
We focus on the latter one, which has isotropic random distribution.%
\footnote{Although there is also an anisotropic random magnetic field component, we neglect 
it~\cite{Haverkorn:2014jka,Boulanger:2018zrk}.}%
We then adopt the smoothly connected cellular model for the distribution of the magnetic field, namely 
there exists homogeneous magnetic fields in a domain whose size is $l_{{\rm G}}$ and
many such cells are contained in the Milky Way Galaxy whose size is $\sim 10$kpc.
The conversion probability between gravitons and photons in a cell can be calculated by diagonalizing Eq.\,(\ref{mat}).
After passing through $N_{{\rm G}} \sim \frac{10{\rm kpc}}{l_{{\rm G}}}$ cells, the total conversion probability may 
be multiplied $N_{{\rm G}}$ times~\cite{Grossman:2002by}, that is
\begin{equation}
  P^{(\sigma)} = N_{{\rm G}}
      \frac{\frac{8 \tilde{B}_{{\rm G}}^{2} \omega^{2}}{M_{{\rm pl}}^{2}}}
      {\left( \omega_{p}^{2} 
            - \omega_{{\rm QED},\sigma}^{2}  -  \omega_{{\rm CMB}}^{2}  \right)^{2}
      + \frac{8 \tilde{B}_{{\rm G}}^{2} \omega^{2}}{M_{{\rm pl}}^{2}} }
      \times
      \sin^{2}\left( \frac{l_{{\rm G}}}{l_{{\rm os}}} \right) , \label{pro}
\end{equation}
where we defined the oscillation length 
\begin{equation}
  l_{{\rm os}} = \frac{4\omega}{\sqrt{\left( \omega_{p}^{2}  - \omega_{{\rm QED},\sigma}^{2}  -  \omega_{{\rm CMB}}^{2} \right)^{2}
                                        + \frac{8 \tilde{B}_{{\rm G}}^{2} \omega^{2}}{M_{{\rm pl}}^{2}} }}  .
\end{equation}
Note that we replaced $B_{{\rm G}}$ by its averaged transverse component $\tilde{B}_{{\rm G}}=\sqrt{2/3}B_{{\rm G}}$.
When the oscillation length $l_{{\rm os}}$ is smaller than the coherence length of magnetic fields $l_{{\rm G}}$,
the sin function in Eq.\,(\ref{pro}) can be replaced by its average, $1/2$, approximately.
Therefore, the conversion probability can be evaluated as follows:
\begin{widetext}
\begin{empheq}[left={P^{(\sigma)} \simeq \empheqlbrace}]{alignat=2}
\ \frac{
  1}{2} \frac{\frac{8 \tilde{B}_{{\rm G}}^{2} \omega^{2}}{M_{{\rm pl}}^{2}}}
      {\left( \omega_{p}^{2} 
            - \omega_{{\rm QED},\sigma}^{2}  -  \omega_{{\rm CMB}}^{2}  \right)^{2}
      + \frac{8 \tilde{B}_{{\rm G}}^{2} \omega^{2}}{M_{{\rm pl}}^{2}} }
 & \quad {\rm for} \quad l_{{\rm os}} < l_{{\rm G}} \label{G1} , \\
\ \frac{N_{{\rm G}}l_{{\rm G}}^{2} \tilde{B}_{{\rm G}}^{2}}{2 M_{pl}^{2}} \hspace{4.5cm}
       & \quad {\rm for} \quad l_{{\rm os}} > l_{{\rm G}}  . \label{G2}
\end{empheq}
\end{widetext}

\begin{figure}[th]
\centering
\includegraphics[width=7.5cm]{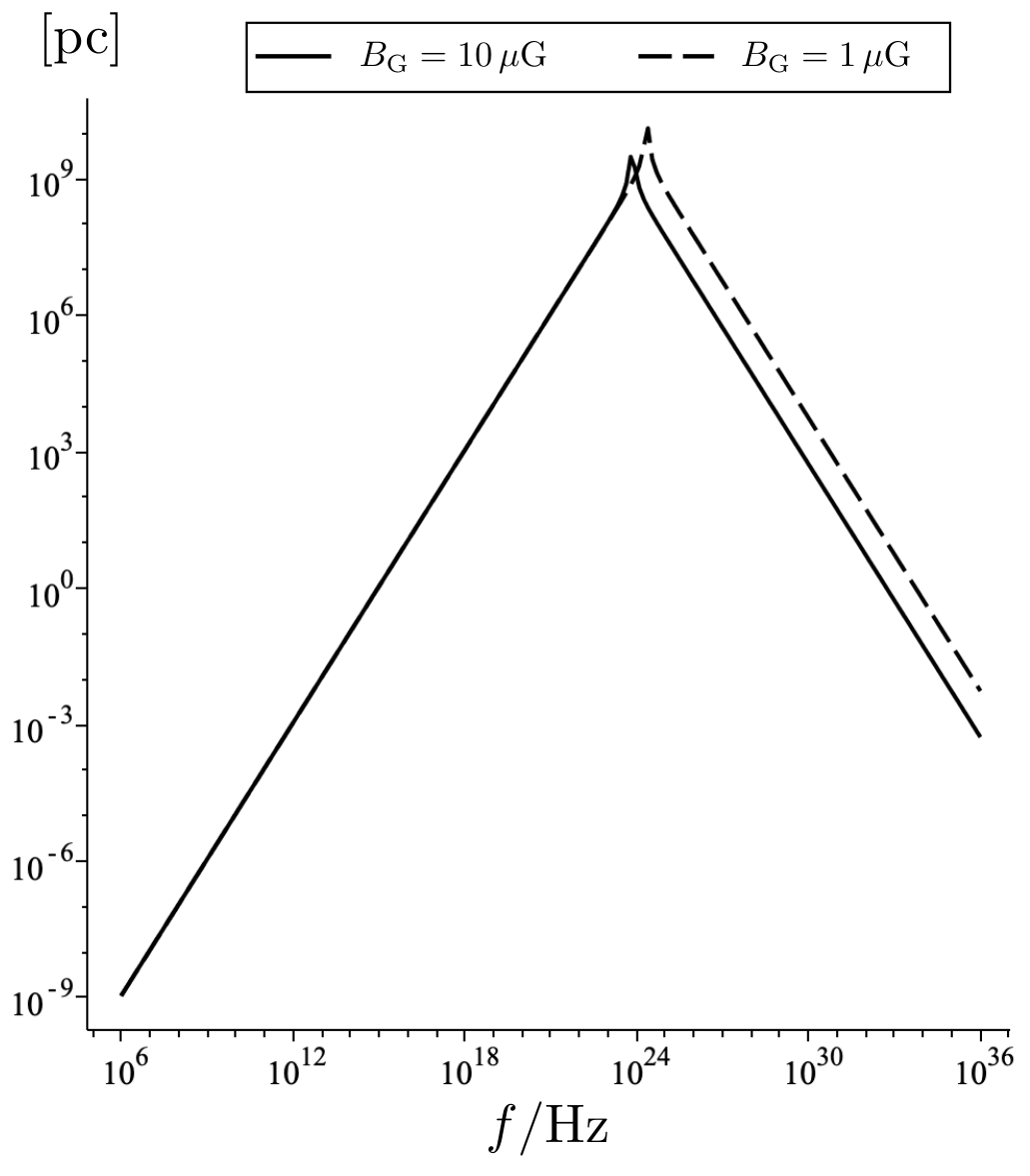}
\caption{Oscillation length within the Milky Way Galaxy ($n_e = 7\times10^{-2} {\rm cm}^{-3}$) 
is shown.
} \label{osc}
\end{figure}

Here is a remark. 
In Eq.\,(\ref{G2}), we implicitly assumed that the direction of the magnetic field suddenly changes at the boundary of the cells, i.e., the variation of the magnetic field is non-adiabatic. 
If $l_{{\rm os}} \gg l_{\rm G}$, this assumption is reasonable.
In the actual situation, however, the transition between two cells may be smooth and hence it would be regarded as adiabatically changing background if $l_{{\rm os}} \ll l_{\rm G}$~\cite{Kartavtsev:2016doq}.
In such a case there is no enhancement by a factor $N_{{\rm G}}$.
Therefore, $N_{{\rm G}}$ has been replaced by $1$ in Eq.\,(\ref{G1}).
The oscillation length is plotted in Fig.~\ref{osc}.
Note that this calculation method is conservative compared to previous works
where the factor $N_{{\rm G}}$ remains even when $l_{{\rm os}} < l_{{\rm G}}$~\cite{Dolgov:2012be}.

Using the above method for calculating the conversion probability,
one can evaluate the photon flux converted from stochastic gravitational waves
in our galaxy with an approximation that the photon flux converted in our galaxy reaches us isotropically.
The electron density varies against the direction and distance from the galactic center~\cite{Cordes:2002wz}.
In the calculation of graviton-photon conversion probability, 
we take a conservative value, $n_e = 7\times10^{-2}  {\rm cm}^{-3}$~\cite{Cordes:2002wz}.
Comparing the calculated photon flux with the observed flux of the CMB~\cite{Hill:2018trh},
CPB~\cite{Hill:2018trh},
active galactic nuclei~\cite{MAGIC:2022piy}, 
the Crab nebula~\cite{LHAASO:2021cbz}, 
and ultra high energy photons~\cite{PierreAuger:2022gkb}, we obtained constraints on stochastic gravitational waves.

\begin{widetext}
\begin{figure}[H]
\centering
\includegraphics[width=16.5cm]{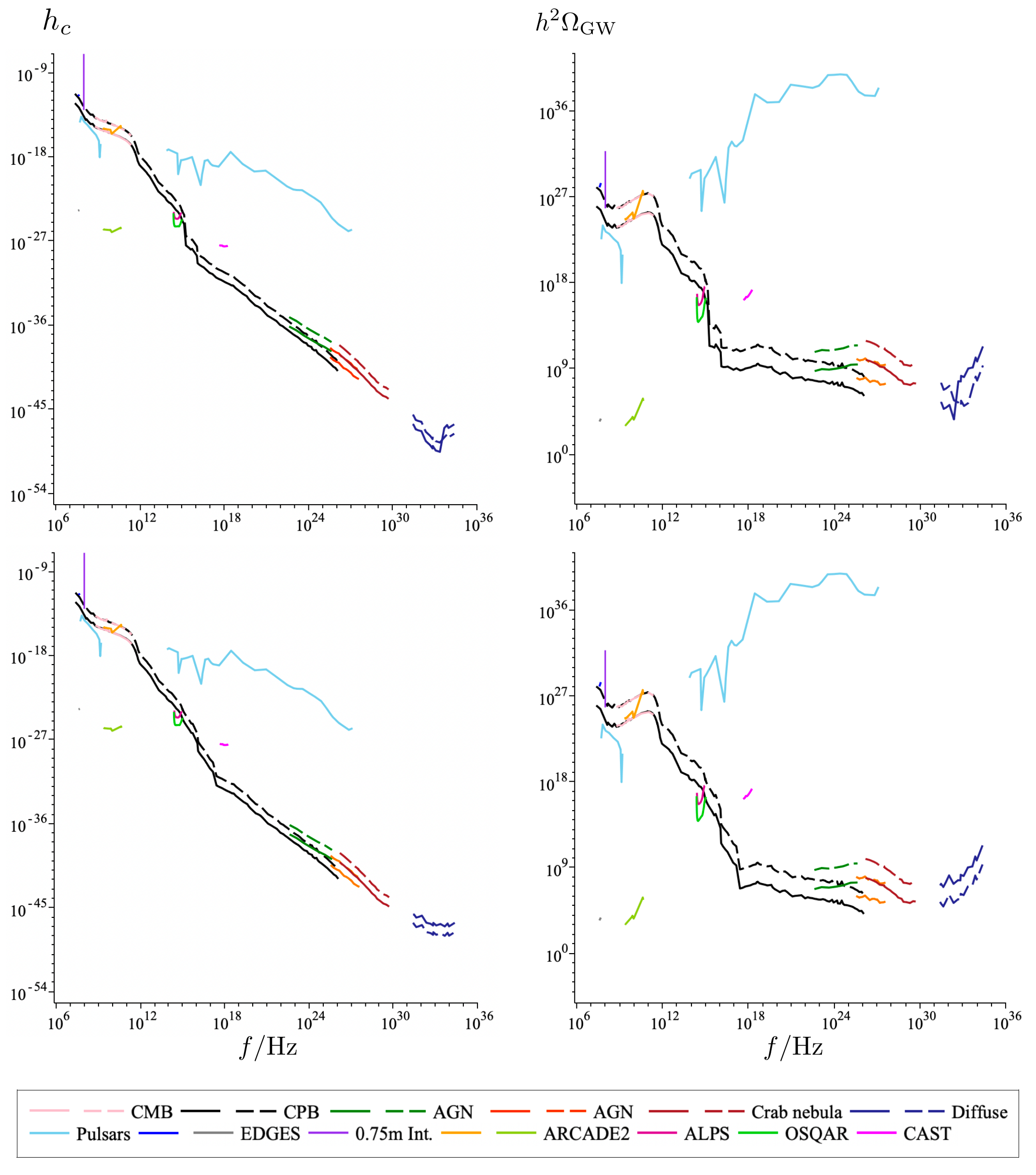}
\caption{Limits on stochastic gravitational waves converted to photons in the Milky Way Galaxy with 
various telescope observations are shown.
The coherence length of magnetic fields is set as
$l_{{\rm G}} = 1$pc for upper figures and $l_{{\rm G}} = 100$pc for lower ones.
The pink, black, deep green, red, brawn, and navy lines (dashed lines) are obtained from 
observations of the CMB~\cite{Hill:2018trh}, the CPB (COBE, AKARI, CIBER, HST, EUVE, MAXI, COMPTEL, FERMI)~\cite{Hill:2018trh},
active galactic nuclei of FERMI~\cite{MAGIC:2022piy} and of MAGIC~\cite{MAGIC:2022piy}, 
the Crab nebula (LHASSO)~\cite{LHAASO:2021cbz}
and ultra high energy photons (Pierre Auger)~\cite{PierreAuger:2022gkb}, respectively, for $B_{{\rm G}}=10\mu$G 
($B_{{\rm G}}=1\mu$G).
The light blue line represents limits from pulasr observations~\cite{Ito:2023fcr}.
The blue (grey) and the orange (lime green) lines respectively represents 
the constraints with EDGES and ARCADE2 for maximal (minimum) amplitude of cosmological magnetic fields~\cite{Domcke:2020yzq}.
The violet line is the upper limit from 0.75~m interferometer~\cite{Akutsu:2008qv}.
The wine red, green, and magenta lines represents constraints with ALPS, OSQAR, and CAST, respectively~\cite{Ejlli:2019bqj}.
} \label{galaxy}
\end{figure}
\end{widetext}

The obtained limits are shown in Fig.\,\ref{galaxy}.
For higher frequencies than $f\sim 10^{33}$Hz for $l_{{\rm G}}=1$pc 
($f\sim 10^{31}$Hz for $l_{{\rm G}}=100$pc), 
one finds that the constraints become stronger for $B_{{\rm G}}=1\mu$G than the case of $B_{{\rm G}}=10\mu$G.
It is because that $\omega_{{\rm QED},\sigma}^4 \propto B_{{\rm G}}^4$ plays an important roll in the denominator 
of Eq.\,(\ref{G1}) when $l_{{\rm os}}<l_{{\rm G}}$.
Thus, the turning points correspond to the frequency when the oscillation length equals the coherent length, i.e.,
$l_{{\rm os}}=l_{{\rm G}}$, as one can see in Fig.\,\ref{osc}.
\subsection{Graviton to photon conversion in intergalactic regions}\label{ingala}
There also exists magnetic fields whose strength $B_{{\rm IG}}$ is considered to range from $10^{-17}$G~\cite{MAGIC:2022piy} to 
$0.1$nG~\cite{Jedamzik:2018itu,Planck:2015zrl,Paoletti:2022gsn,Pshirkov:2015tua}, whose 
coherence length of magnetic fields is $l_{{\rm IG}} \sim 1$Mpc$-4000$Mpc, in intergalactic regions~\cite{MAGIC:2022piy}.
Since there are no known dynamical mechanisms to generate such magnetic fields in the void regions,
they are expected to be produced in the early universe.
We then consider graviton-photon conversion after the photon decoupling to the present time while taking into account
the redshift effect, namely each relevant parameter is promoted to a redshift dependent parameter:
$\omega \rightarrow \omega(1+z)$, $l_{{\rm IG}}\rightarrow l_{{\rm IG}}/(1+z)$, $B_{{\rm IG}}\rightarrow B_{{\rm IG}}(1+z)^{2}$, 
$T\rightarrow T(1+z)$, and $n_{e}\rightarrow n_{b0}(1+z)^{3}\chi_{e}(z)$.
$n_{b0} = 0.25 {\rm m}^{-3}$ is the current baryon number density~\cite{Planck:2018vyg} and 
$\chi_{e}(z)$ is the ionization fraction (we use the model in~\cite{Kunze:2015noa}).
Let us again adopt the smoothly connected cellular model as explained in Sec.~\ref{gala}.
Repeating the similar discussion to Sec.~\ref{gala}, we obtain the conversion probability per time as
\begin{widetext}

\begin{empheq}[left={\tilde{P}^{(\sigma)}(z) \simeq \empheqlbrace}]{alignat=2}
&\ \frac{1}{2l_{{\rm IG}}(z)} \frac{\frac{8 \tilde{B}_{{\rm IG}}^{2}(z) \omega^{2}(z)}{M_{{\rm pl}}^{2}}}
      {\left( \omega_{p}^{2}(z) 
            - \omega_{{\rm QED},\sigma}^{2}(z)  -  \omega_{{\rm CMB}}^{2}(z)  \right)^{2}
      + \frac{8 \tilde{B}_{{\rm IG}}^{2}(z)\omega^{2}(z)}{M_{{\rm pl}}^{2}} }
  &\quad {\rm for}  \quad & l_{{\rm os}}(z) < l_{{\rm IG}}(z) ,  \label{PIG0}   \\
&\ \frac{l_{{\rm IG}}(z) \tilde{B}_{{\rm IG}}^{2}(z)}{2 M_{pl}^{2}}  
       & \quad {\rm for}  \quad & l_{{\rm os}}(z) > l_{{\rm IG}}(z)  .  \label{PIG}
\end{empheq}

\end{widetext}
For $l_{{\rm os}}(z) > l_{{\rm IG}}(z)$, the total conversion rate would be 
\begin{equation}
  \int^{z_{i}(\omega)}_{0} \frac{\tilde{P}^{(\sigma)}(z)}{(1+z) H(z)} dz  ,  \label{IGcon}
\end{equation}
where we used the relation $dt = \frac{-dz}{H(z)(1+z)}$ and  
$H(z) = H_{0} \sqrt{0.69+0.31(1+z)^{3}}$ with $H_{0} = 67{\rm km}/{\rm s}\,{\rm Mpc}$ is the Hubble parameter ~\cite{Planck:2018vyg}.
The integration range is taken from $z=0$ to $z=z_{i}(\omega)$; $z_{i}(\omega)$ represents the time when a photon with a frequency $\omega(z)$ starts to propagate transparently until now.
For example, $z_{i}(\omega)=1100$ for photons with lower frequencies than $\omega(z=1100)<13.6$eV.
However, absorption by atoms can be significant for photons with frequencies higher than $13.6$eV~\cite{Chen:2003gz}.
Moreover, pair creation of an electron and a positron becomes efficient for extremely high energy photons in the presence of CMB or CPB~\cite{Bhattacharjee:1999mup}.
We will set the cutoff of the integration $z_{i}(\omega)$ at the point where the optical depth equals
(approximately) unity~\cite{Chen:2003gz,Bhattacharjee:1999mup}.
The pair creation of an electron and a positron affects not only the absorption but also the dispersion relation of photons.
As discussed in~\cite{Dobrynina:2014qba}, the modified $\omega_{{\rm CMB}}(z)$ can be evaluated with the use of the 
Krasmers-Kronig relation as
\begin{widetext}
\begin{equation}
   \omega_{{\rm CMB}}(\omega,z) \simeq \omega_{{\rm CMB}}(z) \times
          \mathcal{P} \int^{\infty}_{1} 
          \frac{\frac{4x(x+1)-2}{x^{3}} \ln (\sqrt{x}+\sqrt{x-1}) - \frac{2(x+1)\sqrt{x-1}}{x^{5/2}}}{x^{2} - \omega(z)/\omega_{0}},
\end{equation}
\end{widetext}
where the integration denotes the Cauchy principal value. 
$\omega_{0} = \frac{2m_{e}^{2}}{\pi^{4}T_{0}/30 \zeta(3)} = 1.25\times 10^{30}$Hz is the threshold energy for the pair creation.  

Now, as noted in Sec.~\ref{gala}, we must be careful about the use of the cellular model 
for the intergalactic magnetic fields, as each coherent magnetic field may be more smoothly connected.
Firstly, whether $l_{{\rm os}}(z)$ is larger or smaller than $l_{{\rm IG}}(z)$ depends on the redshift. During a period when $l_{{\rm os}}(z) > l_{{\rm IG}}(z)$, the conversion probability per unit time (\ref{PIG}) accumulates. Thus, it should be integrated as shown in Eq.~(\ref{IGcon}).
On the other hand, in a period when $l_{{\rm os}}(z) < l_{{\rm IG}}(z)$,
the total conversion probability would be determined solely by $l_{\rm IG}\tilde{P}^{(\sigma)}$ at 
the point where $l_{{\rm os}}(z) < l_{{\rm IG}}(z)$ breaks.
Therefore, when calculating the total conversion probability, 
we do not integrate $\tilde{P}^{(\sigma)}$ during a period when $l_{{\rm os}}(z) < l_{{\rm IG}}(z)$, 
instead, we add the conversion probability $l_{\rm IG}\tilde{P}^{(\sigma)}$ at the time when
$l_{{\rm os}}(z) < l_{{\rm IG}}(z)$ breaks.
We note that this method generally yields conservative results by neglecting the accumulation effect when 
$l_{{\rm os}}(z) < l_{{\rm IG}}(z)$, 
compared to previous works~\cite{Pshirkov:2009sf, Domcke:2020yzq}.

\begin{widetext}
\begin{figure}[H]
\centering
\includegraphics[width=16.5cm]{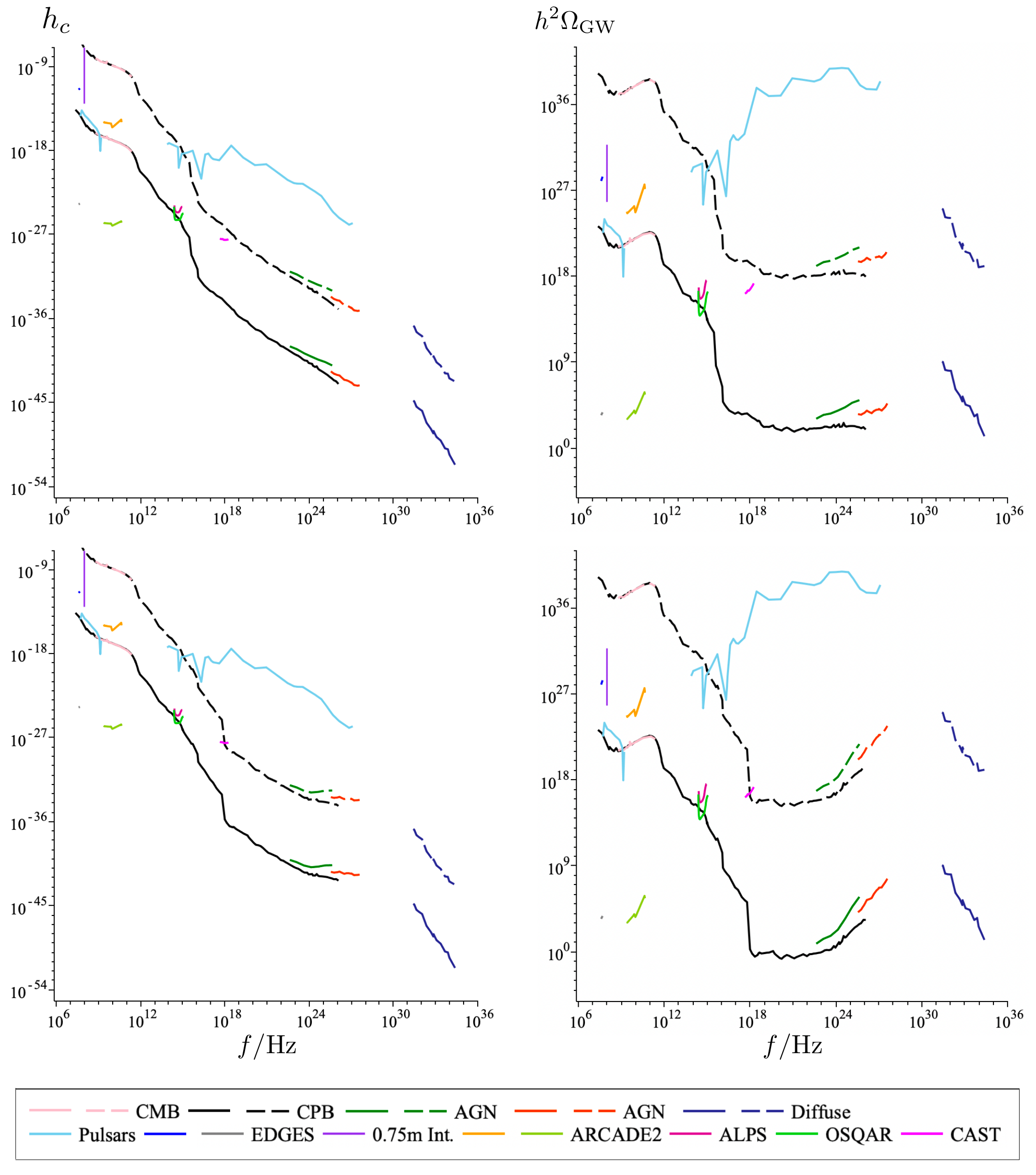}
\caption{Limits on stochastic gravitational waves converted to photons in inter galaxies with various telescope observations are shown.
The coherence length of magnetic fields is set as
$l_{{\rm IG}} = 1$Mpc for upper figures and $l_{{\rm IG}} = 4000$Mpc for lower ones.
The pink, black, deep green, red, and navy lines (dashed lines) are obtained from 
observations of the CMB~\cite{Hill:2018trh}, the CPB (COBE, AKARI, CIBER, HST, EUVE, MAXI, COMPTEL, FERMI)~\cite{Hill:2018trh},
active galactic nuclei of FERMI~\cite{MAGIC:2022piy} and of MAGIC~\cite{MAGIC:2022piy}, 
and ultra high energy photons (Pierre Auger)~\cite{PierreAuger:2022gkb}, respectively, for $B_{{\rm IG}}=0.1$nG 
($B_{{\rm IG}}=10^{-17}$G).
The light blue line represents limits from pulasr observations~\cite{Ito:2023fcr}.
The blue (grey) and the orange (lime green) lines respectively represents 
the constraints with EDGES and ARCADE2 for maximal (minimum) amplitude of cosmological magnetic fields~\cite{Domcke:2020yzq}.
The violet line is the upper limit from 0.75~m interferometer~\cite{Akutsu:2008qv}.
The wine red, green, and magenta lines represents constraints with ALPS, OSQAR, and CAST, respectively~\cite{Ejlli:2019bqj}.
} \label{IG}
\end{figure}
\end{widetext}

Using the above method,
we calculated photon flux from graviton-photon conversion with 
Eqs.\,(\ref{PIG0})-(\ref{IGcon}).
Comparing the calculated photon flux with the observed flux of the CMB~\cite{Hill:2018trh},
CPB~\cite{Hill:2018trh},
active galactic nuclei~\cite{MAGIC:2022piy}, 
and ultra high energy photons~\cite{PierreAuger:2022gkb}, we obtained constraints on the abundance of stochastic gravitational waves.
The obtained limits are shown in Fig.\,\ref{IG}.
The coherence length of magnetic fields is set as
$l_{{\rm IG}} = 1$Mpc for upper figures and $l_{{\rm IG}} = 4000$Mpc for lower ones.
The pink, black, deep green, red, and navy lines (dashed lines) are obtained from 
observations of the CMB~\cite{Hill:2018trh}, the CPB (COBE, AKARI, CIBER, HST, EUVE, MAXI, COMPTEL, FERMI)~\cite{Hill:2018trh},
active galactic nuclei of FERMI~\cite{MAGIC:2022piy} and of MAGIC~\cite{MAGIC:2022piy}, 
and ultra high energy photons (Pierre Auger)~\cite{PierreAuger:2022gkb}, respectively, for $B_{{\rm IG}}=0.1$nG 
($B_{{\rm IG}}=10^{-17}$G).
We note that the observation of PeV gamma-ray~\cite{LHAASO:2021cbz}, which was used in
Sec.~\ref{secearth} and Sec.~\ref{gala}, has not been used here, since
the attenuation effect of PeV gamma-ray by CMB is significant in the extragalactic scale~\cite{Bhattacharjee:1999mup}.
In the low frequency range around GHz, there are also constraints from CMB observations with EDGES and ARCADE2~\cite{Domcke:2020yzq}.
The given constraints are stronger than ours for two reasons: they utilized the information of CMB spectral distortion, and we also employed the smoothly connected cellular model.
Remarkably, one can see that 
for $l_{{\rm IG}} = 4000$Mpc and $B_{{\rm IG}}=0.1$nG, the upper limits on $h^2 \Omega_{{\rm GW}}$ could be less than unity in the frequency range of $10^{18}$-$10^{23}$ Hz. 
\section{Conclusion}
We investigated the graviton-photon conversion under the magnetic fields of the Earth, the Milky Way Galaxy, and intergalactic regions.
To calculate the probability of graviton-photon conversion in the Milky Way Galaxy and in intergalactic regions, we investigated the smoothly connected cellular model that considers the smooth connections among regions containing coherent magnetic fields.
Requiring that the calculated photon flux converted from gravitons does not exceed the observed photon spectra
with various telescopes allows us to constrain the amount of stochastic gravitational waves.
The gravitational wave detection with telescopes enables us to observe high frequency gravitational waves in wide 
frequency ranges above the radio frequency $\sim 10^{7}$Hz.
Remarkably, it pushes the boundaries of frequency in gravitational wave observations to $10^{35}$Hz.
Moreover, the upper limits on $h^2 \Omega_{{\rm GW}}$ could be less than unity in the frequency range of $10^{18}$-$10^{23}$ Hz in a specific case.

It would be desired to prepare templates of photon spectra converted from gravitational wave spectra predicted by each 
source of high frequency gravitational waves~\cite{Ghiglieri:2015nfa,Wang:2019kaf,Ghiglieri:2020mhm,Ringwald:2020ist,Ghiglieri:2022rfp,Vagnozzi:2022qmc,Ema:2021fdz,Ema:2015dka,Ema:2016hlw,Ema:2020ggo,Nakayama:2018ptw,Huang:2019lgd,Barman:2023ymn,Khlebnikov:1997di,Easther:2006gt,Easther:2006vd,
Garcia-Bellido:2007nns,Ito:2016aai,Aggarwal:2020olq,Franciolini:2022htd,Saito:2021sgq,
Gehrman:2022imk,Gehrman:2023esa}
in various background magnetic fields to detect it in data of telescope observations,
as partly demonstrated in this paper.
Although we considered stochastic gravitational waves as a demonstration in this paper, it is also possible to 
observe event-like gravitational waves from such as binaries of light primordial black holes and 
black hole 
superradiance~\cite{Aggarwal:2020olq,Franciolini:2022htd}.
In this case, we have the chance to observe gravitational waves of $h^2 \Omega_{{\rm GW}}\geq 1$.
For example, if primordial black holes with a mass of $\sim 10^{-15}M_{\odot}$ form all of dark matter, we expect 
a merger event from a distance $10$pc per year, which emits a gravitational wave of 
$\sim 10^{19}$Hz at the final stage of the coalescence~\cite{Franciolini:2022htd}.
The photon flux from such a single transient event (not from the accumulated stochastic signals of the merger events) through the graviton-photon conversion in our galaxy is calculated as $\sim 0.001$ erg/${\rm s}\,{\rm cm}^2$, where we assumed 
$B_{{\rm G}} = 10\mu{\rm G}$ and $l_{{\rm G}} = 1$pc.
Apparently, this is detectable with current telescopes.
However, we note that the emission of the gravitational wave only lasts for $\sim 10^{-20}$s~\cite{Franciolini:2022htd}.
This is considerably shorter than the time resolution of current telescopes, which
indicates that further efforts/ideas are needed to push forward detecting gravitational waves with
telescopes.
Nevertheless, we believe that gravitational wave detection with telescopes
offers exciting opportunities to probe new physics which predicts high frequency gravitational 
waves~\cite{Ghiglieri:2015nfa,Wang:2019kaf,Ghiglieri:2020mhm,Ringwald:2020ist,Ghiglieri:2022rfp,Vagnozzi:2022qmc,Ema:2021fdz,Ema:2015dka,Ema:2016hlw,Ema:2020ggo,Nakayama:2018ptw,Huang:2019lgd,Barman:2023ymn,Khlebnikov:1997di,Easther:2006gt,Easther:2006vd,Garcia-Bellido:2007nns,Ito:2016aai,Aggarwal:2020olq,Franciolini:2022htd,Saito:2021sgq,Gehrman:2022imk,Gehrman:2023esa}.
\onecolumngrid
%
\onecolumngrid
%
\onecolumngrid
%
%
%
%
%
%
\begin{acknowledgments}
This work was supported by World Premier International Research Center Initiative (WPI), MEXT, Japan.
A.\,I.\ was in part supported by JSPS KAKENHI Grant Number JP22K14034.
K.\,K.\ was in part supported by MEXT KAKENHI Grant Numbers JP22H05270.
\end{acknowledgments}

\bibliography{draft}

\end{document}